\documentclass[reqno,11pt]{amsart}
\usepackage{amsmath,amssymb,amsthm,graphicx,a4wide,enumerate,url}
\usepackage[small,bf]{caption} 
\usepackage{enumitem,subfigure}
\newcommand{\R}{\mathbb{R}}

\begin{document}
\parskip.9ex

\title[Off-Ramp Coupling Conditions Devoid of Spurious Blocking and Re-Routing]
{Off-Ramp Coupling Conditions Devoid of Spurious Blocking and Re-Routing}
\author[N. Salehi]{Najmeh Salehi}
\address[Najmeh Salehi]
{Department of Mathematics \\ Temple University \\ \newline
1805 North Broad Street \\ Philadelphia, PA 19122}
\email{tuf62199@temple.edu}
\author[J. Somers]{Julia Somers}
\address[Julia Somers]
{Department of Mathematics \\ Temple University \\ \newline
1805 North Broad Street \\ Philadelphia, PA 19122}
\author[B. Seibold]{Benjamin Seibold}
\address[Benjamin Seibold]
{Department of Mathematics \\ Temple University \\ \newline
1805 North Broad Street \\ Philadelphia, PA 19122}
\email{seibold@temple.edu}
\urladdr{http://www.math.temple.edu/\~{}seibold}
\subjclass[2000]{35L65; 35Q91; 91B74}
\keywords{Traffic model, macroscopic, Lighthill-Whitham-Richards, network, coupling condition, off-ramp, vertical queue, split ratio}

\begin{abstract}
When modeling vehicular traffic flow on highway networks via macroscopic models, suitable coupling conditions at the network nodes are crucial. Frequently, the evolution of traffic flow on each network edge is described in a lane-averaged fashion using a single-class Lighthill-Whitham-Richards model. At off-ramps, split ratios (i.e., what percentage of traffic exits the highway) are prescribed that can be drawn from historic data. In this situation, classical FIFO coupling conditions yield unrealistic results, in that a clogged off-ramp yields zero flux through the node. As a remedy, non-FIFO conditions have been proposed. However, as we demonstrate here, those lead to spurious re-routing of vehicles. Then, a new coupling model, FIFOQ, is presented that preserves the desirable properties of non-FIFO models, while not leading to any spurious re-routing.
\end{abstract}

\maketitle

\section{Introduction}
One way to model vehicular traffic flow on a highway network is to describe interchanges and ramps as nodes (vertices) of the network, and the roads between them as edges of the network. The flow on an edge in a given direction is then described via a lane-aggregated Lighthill-Whitham-Richards (LWR) model \cite{LighthillWhitham1955, Richards1956}, and suitable coupling conditions are formulated on each node that describe the flow balance of vehicles between in-going and out-going edges. The classical FIFO coupling conditions were introduced by Daganzo \cite{Daganzo1995_2}. Mathematical proofs of well-posedness of the resulting hyperbolic conservation law network flow were provided in \cite{HoldenRisebro1995, HertyKlar2003, CocliteGaravelloPiccoli2005, GaravelloPiccoli2006} and references therein.

A key modeling shortcoming of FIFO conditions is that a clogged off-ramp will result in zero flow through the node. Clearly, on multi-lane highways this is unrealistic, as a queue forming from the off-ramp will generally be restricted to the right-most lane, and vehicles that do not wish to exit can pass queue (within certain limitations). To remedy this shortcoming, non-FIFO coupling conditions were proposed \cite{lebacque2002first}. These allow for a nonzero flow past a clogged off-ramp, however, at the expense of violating route-choices of drivers: some vehicles that in reality form a queue waiting to exit will instead be accounted for as flow continuing along the highway. Other more recent off-ramp models possess similar re-routing effects \cite{LebacqueLouisSchnetzler2012}.

What is needed are coupling conditions that remedy the shortcoming of FIFO, but without producing spurious re-routing of vehicles. One possible way to achieve this is by formulating multi-commodity models that explicitly track different types of vehicles, such as presented in \cite{bressan2015conservation}. The practical challenge is that (in the present day) it is usually not known in advance which vehicles intend to exit the highway and which do not. Instead, typical split ratios tend to be known from historic data. We therefore consider the situation in which a single-class lane-aggregated LWR model is to be used on the edges, as for instance done in the Mobile Millennium project \cite{MobileMillennium}.

This paper presents how this goal can be achieved by the introduction of a vertical queue that keeps track of the excess vehicles of a certain type (exiting vs.~non-exiting) that may queue up by more than the other vehicle type does. While vertical queues have been used in macroscopic traffic models for the ``storage'' of vehicles that wish to enter the network \cite{delle2014pde, bressan2015conservation}, their use for the purpose of balancing splitting flows is novel. The vertical queue could be present on either of the two out-roads (the highway or the ramp), however, it is always designed to be minimal, i.e., two queues cannot be active simultaneously.

This paper is organized as follows. First, the mathematical definitions and notations are introduced, and a discussion of split ratios vs.~turn ratios is provided. Then, the existing coupling models, FIFO and non-FIFO, are presented, and their modeling shortcomings discussed. After that, the new model is presented, including how it can be implemented in a cell-transmission model (CTM) framework. The three models are then systematically compared in a representative example describing a first forming and then clearing off-ramp queue. The example illustrates how the new model remedies the weaknesses of both existing models. The paper closes with conclusions and future research suggestions.

\vspace{1.5em}
\section{Modeling Foundations}\label{Modeling_Foundations}
We consider models for road networks that are represented by a directed graph, whose edges represent the roads (all lanes going in one direction aggregated), and nodes (vertices) represent the interchanges or ramps (sometimes called ``junctions''). An edge $i$ of the network is an interval $I_i = [a_i,b_i]$. This paper specifically focuses on 1-in-2-out nodes modeling highway off-ramps. Therefore, the discussion is restricted to one node with three edges: one in-road ($I_1$) and two out-roads ($I_2$ and $I_3$).

On each edge $i$, the evolution of the traffic density, $\rho_i(x,t)$, is described by the LWR model
\begin{equation}
\label{eq:LWR}
\partial_t\rho_i+\partial_x f(\rho_i) = 0\;,
\quad\text{where~} (x,t) \in I_i \times \R^+\;.
\end{equation}
The flux function $f = f(\rho) = \rho v(\rho)$ encodes the fundamental diagram (FD) of traffic flow, where $v = v(\rho)$ is the bulk velocity of traffic. In this paper, we use the Greenshields flux
\begin{equation*}
f(\rho) = v^\text{max}\rho\left(1-\frac{\rho}{\rho^\text{max}}\right),
\end{equation*}
corresponding to an affine linear density--velocity relationship, with $v^\text{max}$ the speed limit and $\rho^\text{max}$ the jamming density. However, it is important to stress that the model, its CTM implementation, and its analysis, apply to any concave down flux function, including triangular FDs. The critical density at which the flow is maximized is denoted with $\sigma$.

\subsection{Riemann Problem and Cell Transmission Model}
The key building block for finite volume discretizations of the LWR model \eqref{eq:LWR} is the Riemann problem (RP), which is a Cauchy problem with initial data
\begin{equation*}
\rho(x,0) = \rho_0(x) = \begin{cases}
\rho_\text{L} & x < 0\\
\rho_\text{R} & x \ge 0\;.
\end{cases}
\end{equation*}
By standard theory of hyperbolic conservation laws \cite{LeVeque1992}, the RP (with concave flux) has the following unique entropy solution:
\begin{itemize}
\item If $\rho_\text{L}<\rho_\text{R}$, then $f'(\rho_\text{L})>f'(\rho_\text{R})$, and then the solution
\begin{equation*}
\rho(x,t) = \begin{cases}
\rho_\text{L} & x < st\\
\rho_\text{R} & x \ge st
\end{cases}
\end{equation*}	
consist of a shock, i.e., a traveling discontinuity in which vehicle brake (e.g., the upstream end of a traffic jam). The shock speed $s = \frac{f(\rho_\text{R})-f(\rho_\text{L})}{\rho_\text{R}-\rho_\text{L}}$ is given by the secant slope in the FD.
\item If $\rho_\text{L}\ge\rho_\text{R}$, then $f'(\rho_\text{L})\le f'(\rho_\text{R})$, and the solution
\begin{equation*}
\rho(x,t) = \begin{cases}
\rho_\text{L} & x < f'(\rho_\text{L})t\\
(f')^{-1}(\frac{x}{t}) & f'(\rho_\text{L})t \le x < f'(\rho_\text{R})t\\
\rho_\text{R} & x \ge f'(\rho_\text{R})t
\end{cases}
\end{equation*}
is a rarefaction wave, in which vehicles gradually accelerate.
\end{itemize}
The PR is the key building block of the Godunov method \cite{Godunov1959}, which divides each edge into finite volume cells, and updates the average density in each cell by the numerical fluxes across cell boundaries. Those fluxes are the RP solutions, evaluated at the cell interface. The cell transmission model (CTM) \cite{Daganzo1994} is equivalent to the Godunov method, applied to the LWR model.

An important conceptual interpretation of the Godunov fluxes is in terms of supply and demand functions. Given a concave down flux function $f(\rho)$ with critical density $\sigma$, the supply and demand functions
\begin{equation}
\label{eq:supply_demand_functions}
\gamma^\text{s}(\rho) = \begin{cases}
f(\sigma) & 0 \le \rho < \sigma\\
f(\rho)   & \sigma \le \rho \le 1
\end{cases}
\quad\text{and}\quad
\gamma^\text{d}(\rho) = \begin{cases}
f(\rho)   & 0 \le \rho < \sigma\\
f(\sigma) & \sigma \le \rho \le 1
\end{cases}
\end{equation}
are the non-increasing and non-decreasing components of the flux function, respectively.

The flux of vehicles through an interface between two cells is then the maximum possible value that does not exceed the demand (on road capacity) imposed by the upstream cell (L), and the supply (of road capacity) that the downstream cell (R) provides:
\begin{equation}
\label{eq:Godunov_flux}
F(\rho_\text{L},\rho_\text{R}) = \min\left(\gamma^\text{d}(\rho_\text{L}),\gamma^\text{s}(\rho_\text{R})\right).
\end{equation}
For the implementation of a Godunov scheme, respectively CTM, the interface flux \eqref{eq:Godunov_flux} is all that is needed. However, the full solution of the RP can also be constructed, as follows. On each cell (upstream and downstream), there are two possibilities: if the flux \eqref{eq:Godunov_flux} matches the flux $f(\rho)$ in that cell, then the constant state remains as it is; otherwise, a new state $\hat{\rho}$ emerges at the cell interface that reproduces the interface flux, i.e., $f(\hat{\rho}) = F$. Of the two solutions that this equation generally possesses, the one is chosen that results in a wave that travels away from the cell interface, that is: the congested state $\hat{\rho}>\sigma$ on the upstream cell; and the free-flow state $\hat{\rho}<\sigma$ on the downstream cell.

\subsection{Generalized Riemann Problem}
In a Godunov/CTM discretization of a road network, a network node can be treated in the same fashion. A generalized Riemann Problem (GRP) is given by a constant density state on each edge. For a general node, the demands of all in-roads and the supplies of all out-roads are computed, and by a route choice matrix that determines how the in-fluxes wish to distribute into the out-fluxes, the resulting vehicle flows are constructed so that they never exceed their respective supply/demand values.

In this paper, we focus on the 1-in-2-out case. The GRP considers a constant density $\rho_1$ on the in-road $I_1$ (highway), a constant density $\rho_2$ on the out-road $I_2$ (highway), and a constant density $\rho_3$ on the out-road $I_3$ (off-ramp). A node coupling model (or coupling condition) $\Phi$ is then a mapping from those three densities to three new fluxes $\Phi(\rho_1,\rho_2,\rho_3) = (\Gamma_1,\Gamma_2,\Gamma_3)$, where the flux on the in-road matches the sum of the two out-road fluxes, $\Gamma_1 = \Gamma_2+\Gamma_3$.

Note that by the same construction as in the simple RP, one also obtains three new states $(\hat{\rho}_1,\hat{\rho}_2,\hat{\rho}_3)$ emanating at the node's position on each edge, but as we are interested in Godunov/CTM discretizations, the construction of the fluxes $\Gamma_i$ suffices.

\subsection{Split Ratio}
In a 1-in-2-out node, drivers make route choice decisions. Therefore, realistic coupling models must require an additional piece of information. This is commonly assumed to be a ``split ratio'' that prescribes what ratios of the incoming flow proceeds onto which of the two out-roads. For a general node with multiple in- and out-road, a split ratio matrix is needed \cite{GaravelloPiccoli2006}. In the 1-in-2-out case, the split ratios are given by two numbers $\alpha_2$ and $\alpha_3$, with $\alpha_2+\alpha_3 = 1$, corresponding to the two out-roads $I_2$ and $I_3$, respectively.

While mathematically, the notion of a split ratio (matrix) is easy to accept, its practical/modeling rationale in fact calls for a careful discussion. If both out-roads are in free-flow, then what truly determines how many vehicles exit at a given time is the composition of the incoming traffic flow into ``type 2'' (intending to continue on the highway) and ``type 3'' (intending to exit) vehicles. Unfortunately, this ``type ratio'' is generally not known. (The type may be known for a few vehicles; and future V2X connectivity may substantially increase that knowledge; but for now, the type ratio is not known.) Therefore, a historic ``exit ratio'' is used as a proxy for the unknown ``type ratio''. Assuming that traffic behaves relatively similar from week to week, and assuming that exit ratios evolve slowly (relative to the flow dynamics) in time, historic data on how many vehicles have exited at a certain time of day can be used to define the (quasi-constant-in-time) split ratio $\alpha_2+\alpha_3 = 1$.

The problem with this approach is that the type ratio and the exit ratio are not necessarily the same. As detailed below, they are identical if both out-roads are in free-flow, or if the split happens in a way that passing of other vehicles is impossible. However, at a highway off-ramp, neither of these assumptions needs to be satisfied. As an extreme example, consider an off-ramp that is completely clogged ($\rho_3 = \rho_3^\text{max}$) due to an incident. Hence, no vehicle flow occurs on the ramp, and type 3 vehicles will start to queue up on the highway. However, a multi-lane highway generally allows type 2 vehicles to pass this queue (to some extent). Consequently, the type ratio of vehicles that are upstream of the ramp will change and gradually shift towards more and more type 3 vehicles. However, this is not due to an actual change in upstream traffic, but rather due to the clogged (downstream) off-ramp.

In the following, we present two classical coupling models, that both fail to capture this situation correctly (for different reasons). We then present a new model that remedies the problems.

\vspace{1.5em}
\section{Existing Models}\label{Existing_Models}
As above, we consider a GRP at a 1-in-2-out node, i.e., states $\rho_1$, $\rho_2$, and $\rho_3$ are given. Using equations \eqref{eq:supply_demand_functions}, we obtain the in-road demand $\gamma_1^\text{d}$, and the out-road supplies $\gamma_2^\text{s}$ and $\gamma_3^\text{s}$. Then, using the split ratio $\alpha_2+\alpha_3 = 1$, the ``partial demands'' \cite{jin2003distribution} are given as
\begin{equation*}
\gamma_{21}^\text{d} = \alpha_2\gamma_{1}^\text{d}
\quad\text{and}\quad
\gamma_{31}^\text{d} = \alpha_3\gamma_{1}^\text{d}\;.
\end{equation*}
Below we always assume that traffic actually ``splits'', i.e., $\alpha_2>0$ and $\alpha_3>0$.

\subsection{FIFO Model}\label{FIFO_Model}
The FIFO (``first-in-first-out'') coupling model \cite{Daganzo1995_2} is based on the assumption that the actual exit ratio equals the split ratio under all circumstances. Hence, the new fluxes satisfy
\begin{equation}
\label{eq:FIFO_conditions}
\Gamma_2 = \alpha_2\Gamma_1
\quad\text{and}\quad
\Gamma_3 = \alpha_3\Gamma_1\;.
\end{equation}
As described above, the GRP requires that the new fluxes do not exceed the respective supplies/demands on the edges, i.e., $0\le \Gamma_1 \le \gamma_1^\text{d}$,~~$0\le \Gamma_2 \le \gamma_2^\text{s}$, and $0\le \Gamma_3 \le \gamma_3^\text{s}$. Using \eqref{eq:FIFO_conditions}, the latter two conditions can be re-written as $0\le \Gamma_1 \le \frac{1}{\alpha_2}\gamma_2^\text{s}$ and $0\le \Gamma_1 \le \frac{1}{\alpha_3}\gamma_3^\text{s}$. Maximizing the flux through the node under those constraints determines the fluxes as
\begin{equation*}
\Gamma_1 = \min\left(\gamma_1^\text{d}, \frac{\gamma_2^\text{s}}{\alpha_2}, \frac{\gamma_3^\text{s}}{\alpha_3}\right),\quad
\Gamma_2 = \alpha_2\Gamma_1\;,\quad
\Gamma_3 = \alpha_3\Gamma_1\;.
\end{equation*}
The FIFO model ensures that the resulting fluxes are always distributed according to the prescribed split ratio. Therefore, in the case of a clogged off-ramp $I_3$ (but free-flow $I_2$), FIFO would result in zero flow through the node, $\Gamma_1 = \Gamma_2 = \Gamma_3 = 0$. Clearly, this is unrealistic for highway off-ramps, whose multiple lanes allow vehicles to by-pass queues (to some extent), and vehicles waiting to pass through the node do not necessarily hold up all traffic. In other words, highways are clearly not FIFO.

\subsection{Non-FIFO Model}\label{non_FIFO_Model}
Using partial demands, the out-fluxes in FIFO can equivalently be written as
\begin{equation*}
\Gamma_j = \min\left(\gamma_{j1}^\text{d}, \frac{\alpha_j}{\alpha_2}\gamma_2^\text{s}, \frac{\alpha_j}{\alpha_3}\gamma_3^\text{s}\right),\quad j = 2,3\;.
\end{equation*}
The idea of the non-FIFO model, proposed in \cite{lebacque2002first}, is to associate the supply constraints of each out-road only with the flux on that respective road, leading to the model
\begin{equation*}
\Gamma_2 = \min(\gamma_{21}^\text{d},\gamma_2^\text{s})\;,\quad
\Gamma_3 = \min(\gamma_{31}^\text{d},\gamma_3^\text{s})\;,\quad
\Gamma_1 = \Gamma_2+\Gamma_3\;.
\end{equation*}
A physical interpretation of this model is that, before reaching the node, drivers are already presorted according to the respective out-road that they plan to take. Then, respective portions of the road width are allocated to the drivers according to the split ratio. Thus, each type of driver can pursue their destination without impediment from the other type.

By construction, the non-FIFO model does not incur the blockage problem that FIFO incurs. In the case of a clogged off-ramp (i.e., $\gamma_3^\text{s} = 0$), one has $\Gamma_3 = 0$ (as it has to be), but $\Gamma_2 > 0$ in general. Type 2 vehicles can pass the off-ramp queue, and the flux $\Gamma_2$ is determined solely by the demand and supply of the highway segments.

Unfortunately, the non-FIFO model suffers from a different modeling problem. It assumes that the split ratios are given and independent of the actual density states and fluxes. However, because the out-fluxes are generally \emph{not} distributed according to the split ratio (i.e., $\frac{\Gamma_2}{\Gamma_3} \neq \frac{\alpha_2}{\alpha_3}$), vehicles of one type will actually become more prevalent upstream of the node than vehicles of the other type. The non-FIFO model ignores this fact: once the clogged off-ramp becomes free-flow again, vehicles of type 3 will start flowing again; however, the fact that a queue of type 3 vehicles should be present is ignored. This results in a spurious re-routing of vehicles: vehicles that should have taken the off-ramp will be allocated to out-flow on the highway instead.

To recap, the non-FIFO model remedies the unrealistic blockage imposed by the FIFO model. However, it does so at the expense of unphysical re-routing effects. It should be stressed that under certain circumstances, drivers in reality may in fact change their route choices based on the actual traffic state; however, this is not always possible, and a model should not produce re-routing as an unwanted side effect. To describe intentional route changing, models that allow the split ratios to depend on the traffic state have been proposed \cite{herty2003modeling}. However, these models do not remove the fundamental flaws of FIFO and non-FIFO. A methodology that would resolve the re-routing problem is to explicitly track and evolve the split ratios (or more accurately: the type ratios) as they move along the incoming edge, as proposed for instance by Bressan and Nguyen \cite{bressan2015conservation}. In addition to being computationally substantially more demanding, such explicit multi-commodity models are held back by the aforementioned fundamental challenge that the route choices of vehicles are generally not accessible in advance.

The question is therefore, what can be done to remedy the problems of both models (FIFO and non-FIFO) within the framework of single-class, lane-aggregated, macroscopic models. The new model should still be based on historic split ratios; it should allow for passing of queued vehicles; it should not lose any vehicles (that are waiting in queues); and it should account for the type of queued vehicles. As the formation of a queue, caused by reduced supply of one out-road, introduces non-local-in-time effects (vehicles queued up initially may still wait later in time), imbalances in the vehicle type composition among queued vehicles must, in some form, be tracked. Next, we construct a model that does so, with the minimal amount of additional information stored, namely the excess of vehicles of a certain type, relative to the actual split ratio.

\vspace{1.5em}
\section{New FIFOQ Model}
We now derive a new model that remedies both shortcomings of FIFO and non-FIFO, called FIFOQ (``FIFO with Queue''). As argued above, the proper evolution of waiting and passing vehicles requires the model to be augmented by some additional variable that accounts for the composition of backed-up vehicle types. We choose to introduce a local vertical queue as that additional variable. Because vertical queues improperly capture the non-local impact of true vehicle jamming, we derive a model that minimizes the impact of the vertical queue, so that its sole purpose is to account for excess vehicles of a certain type. It should be stressed that the use for vertical queues in traffic models is not novel at all. For instance, vertical queues are a common means to implement vehicles that enter the network \cite{delle2014pde, bressan2015conservation}. In contrast, the usage of queues for the purpose of properly tracking vehicle types upstream of an off-ramp is a novel concept.

The fundamental dynamics of a vertical queue are that its rate of change equals inflow into its upstream end, $\Gamma_\text{in}$, minus outflow out of its downstream front, $\Gamma_\text{out}$, i.e., $\dot{m} = \Gamma_\text{in}-\Gamma_\text{out}$, where $m$ denotes the number of vehicles stored in the queue \cite{vickrey1969congestion}.

\begin{figure}[h]
\centering
\includegraphics[scale=1]{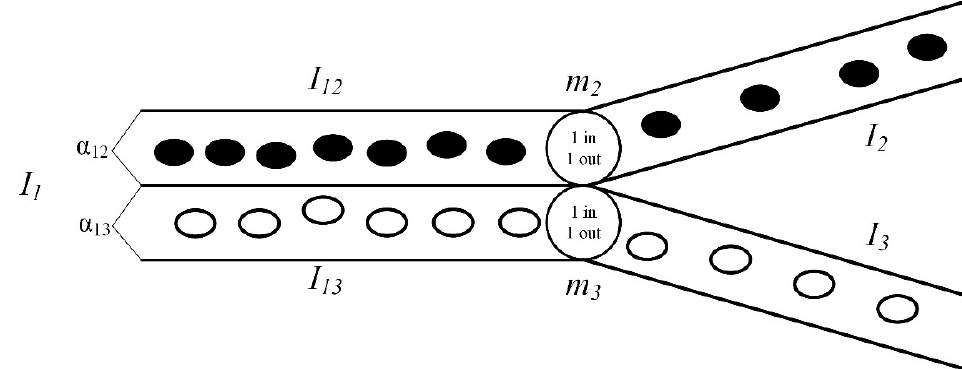}	
\caption{A non-FIFO model with queues.}
\label{fig:separated_split}
\end{figure}

To derive the new model, we start with a preliminary setup: a non-FIFO model with queues. Recall that the non-FIFO model can be interpreted as traffic being presorted and vehicle types given a share of the in-road width according to the split ratio. We therefore think of separating the in-road into two parts and consider two independent 1-in-1-out nodes with queues $m_2$ and $m_3$, see Figure~\ref{fig:separated_split}. These queues could now be used, for example, to prevent backward going shocks on the in-road. However, this is not an adequate model because it does not guarantee a preservation of the split ratio. Moreover, two queues may form simultaneously.

\begin{figure}[h]
\centering	
\includegraphics[scale=1]{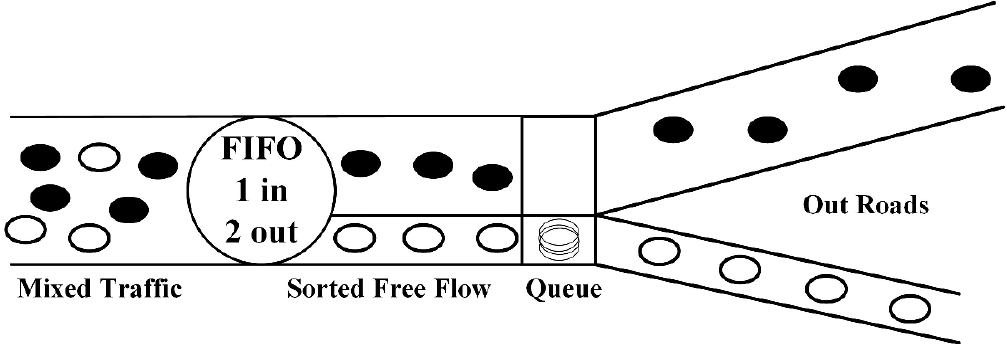}	
\caption{New model: 1-in-2-out FIFO node with Queue (``FIFOQ'').}
\label{fig:full_model}
\end{figure}

We therefore extend the model by adding additional components. As shown in Figure~\ref{fig:full_model}, a 1-in-2-out FIFO node serves to sort the vehicles into two groups based on their destinations. Next downstream, a free flow section follows, which consists of two independent pipes with flux functions $f_{1i} = v_1^\text{max} \rho(1-\frac{\rho}{\rho_1^\text{max}c_i})$ for $i=2,3$, where $c_2$ and $c_3$ are the road sharing ratios of different vehicle types. In many situations, one highway lane will be associated with the off-ramp, while the remaining lanes are associated with the flow past the off-ramp traffic, and these geometric considerations could be incorporated into the model. In this paper, however, we use a simplifying assumption, namely that the road is divided precisely according to the split ratio, i.e., $c_2 = \alpha_2$ and $c_3 = \alpha_3$. The reason why we can assume this two-pipe region to be in free-flow is that the two queues, $m_2$ and $m_3$, can be used to absorb any congestion that emanates from the two out-roads. Of course, these individual components are for model derivation purposes only; in the end, the whole model is ``collapsed to zero length'' to yield a single coupling condition.

Thanks to the queues, this model allows for traffic to pass through the junction while still respecting the split ratio, even when one out-road is clogged. However, the model can further be improved. In the case when both out-roads are clogged (or provide sufficiently low supply), this current model would develop two queues, one in each pipe. In turn, it would never create a congested state on the in-road. That is unsatisfactory, as the impact that a real traffic jam would have on the road conditions further upstream would not be seen. We therefore modify the model to allow for backward propagating shocks on the in-road, as long as that congested state is composed according to the split ratio. We therefore introduce a quantity $\mu$ that is removed from the flux into the node, so that at most one queue is active at any instant in time. That one active queue then tracks the excess of vehicles of a certain type that are more prevalent near the off-ramp than the split-ratio would dictate.

Let $\tilde{\Gamma}_1$, $\tilde{\Gamma}_{12}$, $\tilde{\Gamma}_{13}$, $\tilde{m}_2$, and $\tilde{m}_3$, denote the fluxes and queue values before the removal of $\mu$. We assume that the two-pipe region is in free flow initially. Thus, the supplies of the pipes, $\gamma_{12}^\text{s}$ and $\gamma_{13}^\text{s}$, are always maximal.

Because we here assume that traffic divides into two pipes whose width ratio equals the split ratio, we simply have $\tilde{\Gamma}_1 = \gamma_1^\text{d}$ (if the pipe widths were different, $\tilde{\Gamma}_1 < \gamma_1^\text{d}$ could occur). Moreover, $\tilde{\Gamma}_{12} = \alpha_2\tilde{\Gamma}_1=\alpha_2\gamma_1^\text{d}$ and $\tilde{\Gamma}_{13} = \alpha_3\tilde{\Gamma}_1=\alpha_3\gamma_1^\text{d}$.

Collapsing the pipes to have length implies that the flux out of them equals the flux into them. Therefore, the fluxes into the 1-in-1-out nodes (or queues) are $\tilde{\Gamma}_{12}$ and $\tilde{\Gamma}_{13}$, respectively. If there are no queues, the out-road fluxes are
\begin{equation}
\label{s1}
\Gamma_2 = \min\left(\alpha_2\gamma_1^\text{d}, \gamma_2^\text{s}\right)
\quad\text{and}\quad
\Gamma_3 = \min\left(\alpha_3\gamma_1^\text{d}, \gamma_3^\text{s}\right).
\end{equation}
In turn, if a queue is active, i.e., $\tilde{m}_i > 0$, then the corresponding out-flux is maximal, i.e., $\Gamma_i = \gamma_i^\text{s}$. The evolution of the queues is described by the differences in the fluxes: $\dot{\tilde{m}}_2 = \tilde{\Gamma}_{12}-\Gamma_2 = \alpha_2\tilde{\Gamma}_1-\Gamma_2$ and $\dot{\tilde{m}}_3 = \tilde{\Gamma}_{13}-\Gamma_3 = \alpha_3\tilde{\Gamma}_1-\Gamma_3$.

As by the prior discussion, we now determine $\mu$ such that at most one queue is active at any time. If one queue is active, for example $\tilde{m}_2 > 0$, and the other queue is filling, i.e., $\dot{\tilde{m}}_3 > 0$, then $\mu = \frac{\dot{\tilde{m}}_3}{\alpha_3}$. With $\Gamma_1 = \gamma_1^\text{d}-\mu$, we obtain for the in-flux
\begin{equation*}
\Gamma_1 = \min\left(\gamma_1^\text{d}, \frac{\gamma_3^\text{s}}{\alpha_3}\right).
\end{equation*}
Conversely, if no queue is active, i.e., $\tilde{m}_2 = \tilde{m}_3 = 0$, we have
\begin{equation}
\label{s2}
\mu = \min\left(\frac{\dot{\tilde{m}}_2}{\alpha_2},\frac{\dot{\tilde{m}}_3}{\alpha_3}\right).
\end{equation}
By combining equations~\eqref{s1} and~\eqref{s2}, we have
\begin{equation*}
\mu
= \min\left(\gamma_1^\text{d}-\min\left(\gamma_1^\text{d},\frac{\gamma_2^\text{s}}{\alpha_2}\right),\gamma_1^\text{d}-\min\left(\gamma_1^\text{d},\frac{\gamma_3^\text{s}}{\alpha_3}\right)\right)
= \gamma_1^\text{d}-\min\left(\gamma_1^\text{d},\max\left(\frac{\gamma_2^\text{s}}{\alpha_2},\frac{\gamma_3^\text{s}}{\alpha_3}\right)\right),
\end{equation*}
and with $\Gamma_1 = \gamma_1^\text{d}-\mu$, we obtain for the in-flux
\begin{equation*}
\Gamma_1 = \min\left(\gamma_1^\text{d}, \max\left(\frac{\gamma_2^\text{s}}{\alpha_2},\frac{\gamma_3^\text{s}}{\alpha_3}\right)\right).
\end{equation*}
Therefore the complete model reads as:\vspace{-.5em}
\begin{align*}
&\text{If~} m_2 = m_3 = 0:
\begin{cases}
\Gamma_1 = \min\left(\gamma_1^\text{d}, \max\left(\frac{\gamma_2^\text{s}}{\alpha_2},\frac{\gamma_3^\text{s}}{\alpha_3}\right)\right)\\
\Gamma_2 = \min\left(\alpha_2\gamma_1^\text{d}, \gamma_2^\text{s}\right)\\
\Gamma_3 = \min\left(\alpha_3\gamma_1^\text{d}, \gamma_3^\text{s}\right)
\end{cases}\\
&\text{If~} m_2 > 0:\hspace{2.4em}
\begin{cases}
\Gamma_1 = \min\left(\gamma_1^\text{d}, \frac{\gamma_3^\text{s}}{\alpha_3}\right)\\
\Gamma_2 = \gamma_2^\text{s}\\
\Gamma_3 = \min\left(\alpha_3\gamma_1^\text{d},\gamma_3^\text{s}\right)
\end{cases}\\
&\text{If~} m_3 > 0:\hspace{2.4em}
\begin{cases}
\Gamma_1 = \min\left(\gamma_1^\text{d}, \frac{\gamma_2^\text{s}}{\alpha_2}\right)\\
\Gamma_2 = \min\left(\alpha_2\gamma_1^\text{d},\gamma_2^\text{s}\right)\\
\Gamma_3 = \gamma_3^\text{s}
\end{cases}\\
&\dot{m}_2 = \alpha_2\Gamma_1-\Gamma_2\\
&\dot{m}_3 = \alpha_3\Gamma_1-\Gamma_3
\end{align*}
By construction, this model never generates more than one queue to be active. If, for example, $m_3>0$, then the model dynamics automatically imply that $\dot{m}_2 = 0$. Moreover, in the case of a singular split ratio, such as $\alpha_3 = 0$, we let $\mu = \frac{\dot{\tilde{m}}_2}{\alpha_2}$, in which case the model reduces to the standard FIFO model for a 1-in-1-out node.

\vspace{1.5em}
\section{Cell Transmission Discretization}
We now describe how the new model is discretized into a CTM, by suitably augmenting the Godunov scheme \cite{Godunov1959} with a treatment of the queue evolution.

\subsection{Approximation Along Edges}
Let space and time be discretized via a regular grid, where
\begin{enumerate}[label=(\roman*)]
\item $\Delta x$ is the cell size;
\item $\Delta t$ is the time step, adhering to the CFL conditions $\Delta t \le \frac{\Delta x}{\max |f'(\rho)|}$, where the maximum is taken over all flux functions in the network; and
\item $(x_j,t^n) = (j\Delta x,n\Delta t)$ are the space-time grid points, and $\rho_j^n$ denotes the (average) density on cell $j$ at time $t_n$.
\end{enumerate}
Using $\lambda = \frac{\Delta t}{\Delta x}$, the Godunov scheme along an edge reads as
\begin{equation}
\label{eq:godunov}
\rho^{n+1}_{j} = \rho^{n}_{j} - \lambda (F(\rho^{n}_{j+1},\rho^{n}_{j})-F(\rho^{n}_{j-1},\rho^{n}_{j}))\;,
\end{equation}
where, for internal cell boundaries, the numerical flux
\begin{equation*}
F(\rho_\text{L}, \rho_\text{R}) = \begin{cases}
\min_{\rho_\text{L}\le \rho \le \rho_\text{R}}  f(\rho) \quad &\text{if~} \rho_\text{L} \le \rho_\text{R}\\
\max_{\rho_\text{R}\le \rho \le \rho_\text{L}} f(\rho) \quad &\text{if~} \rho_\text{R} \le \rho_\text{L}
\end{cases}
\end{equation*}
equals the standard CTM flux. At terminal cells of an edge, i.e., adjacent to a network node, the in-flux or out-flux in equation \eqref{eq:godunov} are replaced by the respective flux $\Gamma_i$ defined by the coupling model. Moreover, at boundaries that are not connected to a node, ghost cells are used (cf.~\cite{CocliteGaravelloPiccoli2005}).

\subsection{Treatment of Queues}
The proper time-stepping of the model with queues must take into account that a queue may deplete during a time step, see \cite{delle2014pde}. For simplicity of notation, we describe the situation of a 1-in-1-out node with a vertical queue $m$. At each time, we must determine the new length of the queue. If the queue is filling, the increment is simply added to $m^n$. However, if the queue is emptying, we must calculate the time of queue depletion, $\bar{t} = t^n+\frac{m^n}{\Gamma_2-\Gamma_1}$, and compare it with the time $t^{n+1} = t^n+\Delta t$. We have the following cases:
\begin{itemize}
\item If $\Gamma_1 \ge \Gamma_2$, then: $m^{n+1} = m^n+\Delta t (\Gamma_1-\Gamma_2)$.
\item If $\Gamma_1 < \Gamma_2$, then: $m^{n+1} = \begin{cases}
m^n+\Delta t (\Gamma_1-\Gamma_2) &\text{if~} \Delta t \le \bar{t}-t^n\\
0 &\text{if~} \Delta t > \bar{t}-t^n\;.
\end{cases}$
\end{itemize}
We now show the important property that if a queue empties in a given time step, it remains empty until the end of the step. If $\dot{m} < 0$, the GRP at the junction has a switching point when $m = 0$. Thus, we consider not only the initial states $\rho_1$, $\rho_2$ and the final states $\hat{\rho}_1$, $\hat{\rho}_2$, but also intermediate states $\bar{\rho}_1$, $\bar{\rho}_2$, i.e. the states at $\bar{t}$. While the queue is still emptying, we have $\Gamma_1 = f(\rho_1)$ and $\Gamma_2 = \gamma^\text{s}_2(\rho_2)$. Then, at time $\bar{t}$, when $m = 0$, we consider the GRP with initial states $\bar{\rho}_1$ and $\bar{\rho}_2$. Since by the definition of the demand function we have: $ f(\bar{\rho}_1) \le \gamma^\text{d}_1(\bar{\rho}_1)$, and because the queue was depleting, we will have $f(\bar{\rho}_1) < \gamma^\text{s}_2(\bar{\rho}_2)$. Therefore, we have that $f(\bar{\rho}_1) \le \min(\gamma^\text{d}_1(\bar{\rho}_1),\gamma^\text{s}_2(\bar{\rho}_2))$. Thus, the queue does not begin to fill again.

We also need to modify the Godunov scheme in the case of an emptying queue. We divide the time step into two sub-intervals, $(t^n,\bar{t})$ and $(\bar{t},t^{n+1})$, where $\Delta t_a = \bar{t}-t^n$ and $\Delta t_b = t^{n+1}-\bar{t}$. Then, we solve two different RPs. For $\Delta t_a$, we solve the classical Godunov scheme, and for $\Delta t_b$, we solve another RP with fluxes as given by the case when $m = 0$. Thus, the total fluxes over the full time step add up to
\begin{align*}
\Gamma_1 &= \frac{\Delta t_a}{\Delta t}f_1(\rho_1)
+\frac{\Delta t_b}{\Delta t}\min(\gamma^\text{d}_1(\bar{\rho}_1),\gamma^\text{s}_2(\bar{\rho}_2))\;,\\
\Gamma_2 &= \frac{\Delta t_a}{\Delta t}\gamma^\text{s}_2({\rho}_2)
+\frac{\Delta t_b}{\Delta t}\min(\gamma^\text{d}_1(\bar{\rho}_1),\gamma^\text{s}_2(\bar{\rho}_2))\;.
\end{align*}

\vspace{1.5em}
\section{Model Comparison in a Representative Example} \label{Model_Comparison_Example}
Here we construct a specific example that highlights the key differences of the three models, FIFO, non-FIFO, and FIFO with queue (FIFOQ). In this example, the off-ramp is set to be completely clogged at the initial time ($t=0$), and after some fixed time it is (artificially) set to free-flow. Then, at the final time, the total tally of vehicles exiting and passing the off-ramp is taken.

We use the Greenshields flux function with a critical density of 40 veh/km/lane, and a maximum velocity of 100 km/h, which yields a capacity of 2000 veh/h/lane. For the three edges, we use
\begin{align*}
&\rho_1^\text{max} = \rho_2^\text{max} = 320\text{~veh/km}
\quad\text{and}\quad
\rho_3^\text{max} = 80\text{~veh/km}\;,\\
&v_1^\text{max} = v_2^\text{max} = v_3^\text{max} = 100\text{~km/h}\;,
\end{align*}
representing a 4-lane highway and a single-lane off-ramp (for simplicity with the same speed limit as the highway). Moreover, a split ratio of $\alpha_2=\frac{5}{6}$ and $\alpha_3=\frac{1}{6}$ is assumed, and the initial densities are
$\rho_1 = 0.4\,\rho_1^\text{max}$, $\rho_2 = 0$, and $\rho_3 = \rho_3^\text{max}$, representing a free-flow highway with a completely clogged off-ramp. This situation is let to evolve until $t=9$ min. Then, we perform an idealized maneuver (for the sake of simplicity) in which we set $\rho_3 = 0$, i.e., we instantaneously remove all vehicles from the off-ramp. We then let this new situation evolve until $t = 25$ min. At that final time, we tally the total (time-integrated) fluxes $I_2$ and $I_3$ onto the highway and the off-ramp, respectively.

\begin{table}[tb]
\centering
\begin{tabular}{|c|cccc|}
\hline
Model & Total in-flux & Total out-flux on $I_2$ & Total off-ramp flux & Ratio of out-fluxes \\ \hline
FIFO            & \phantom{1}8000 veh & \phantom{1}6667 veh & 1333 veh & 5:1    \\
non-FIFO        & 11750 veh & 10417 veh & 1333 veh & 7.81:1 \\
FIFOQ           & 12000 veh & 10000 veh & 2000 veh & 5:1    \\
\hline
\end{tabular}
\vspace{.2em}
\caption{Total vehicle flow after $t=25$ min.}
\label{tab:total_fluxes}
\end{table}

Table~\ref{tab:total_fluxes} shows the total vehicle fluxes, obtained by the different models. At the end of this experiment, the queue that forms in the FIFOQ model has emptied. One can clearly see the key shortcomings of the existing models. The FIFO model respects split ratios, but it predicts zero total flux up until $t=9$ min, resulting in severely reduced highway flux. In turn, the non-FIFO model produces reasonable highway flows, but it fails to respect the split ratio: due to the negligence of the off-ramp queue that arises in reality, it falsely re-allocates vehicles that intended to exit, into highway flow.

The new FIFOQ model remedies both of those shortcomings. Unlike FIFO, it produces a nonzero highway flux before $t=9$ min. However, unlike non-FIFO, it tracks the accumulation of type 3 vehicles via the growing vertical queue $m_3$. Then, after $t=9$, this queue depletes, releasing those queued type 3 vehicles onto the off-ramp. Therefore, the total off-ramp flux produced by the FIFOQ model is noticeably larger than with the other two models; which is more realistic, because it remedies the spurious re-routing of the non-FIFO model.

\newlength{\subfigwidth}\setlength{\subfigwidth}{0.815\textwidth}

\begin{figure}[p]
\centering
\subfigure[$t=1.5$ min]{\parbox{\subfigwidth}{\label{fig:FIFO_t_1_5}\includegraphics[width=\subfigwidth]{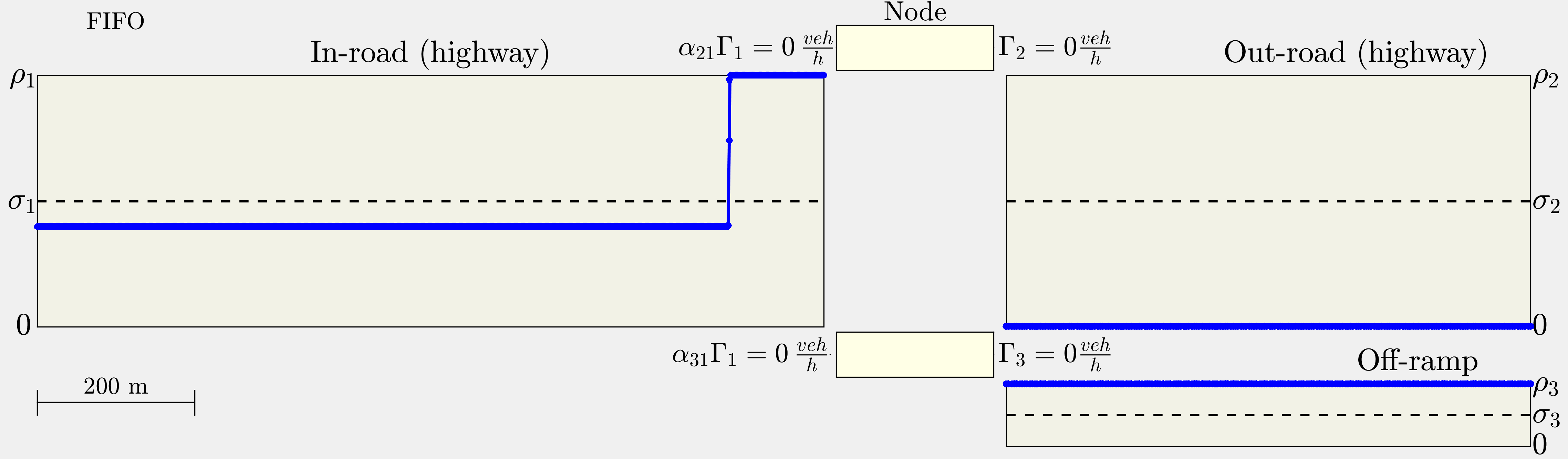}}}\\[-.1em]
\subfigure[$t=6$ min]{\parbox{\subfigwidth}{\label{fig:FIFO_t_6}\includegraphics[width=\subfigwidth]{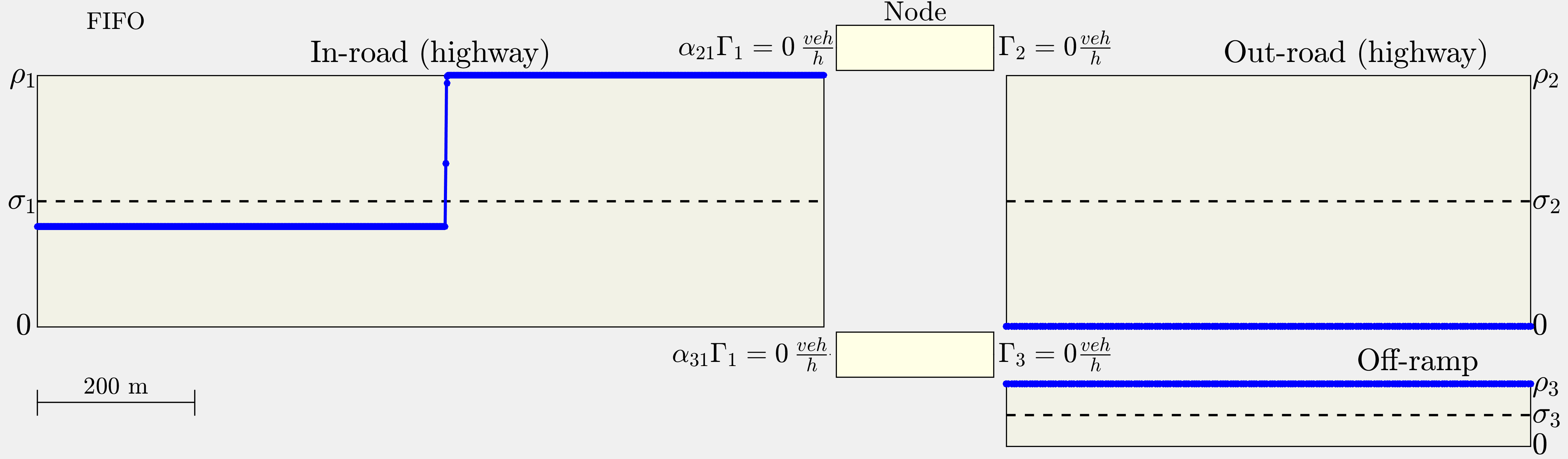}}}\\[-.1em]
\subfigure[$t=9.5$ min]{\parbox{\subfigwidth}{\label{fig:FIFO_t_9_5}\includegraphics[width=\subfigwidth]{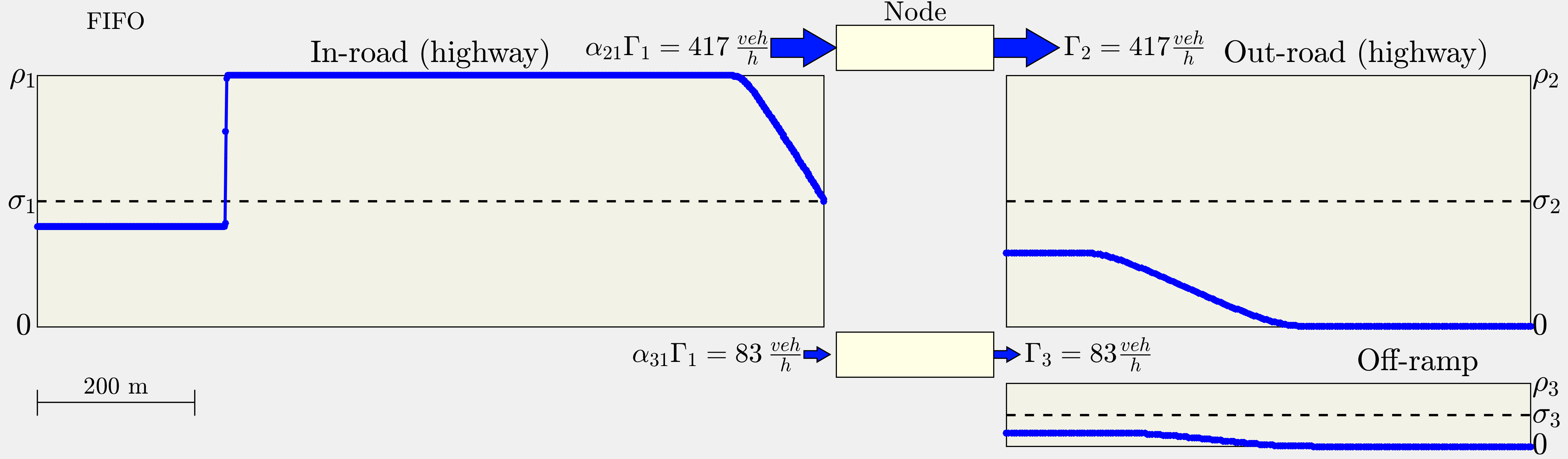}}}\\[-.1em]
\subfigure[$t=15$ min]{\parbox{\subfigwidth}{\label{fig:FIFO_t_15}\includegraphics[width=\subfigwidth]{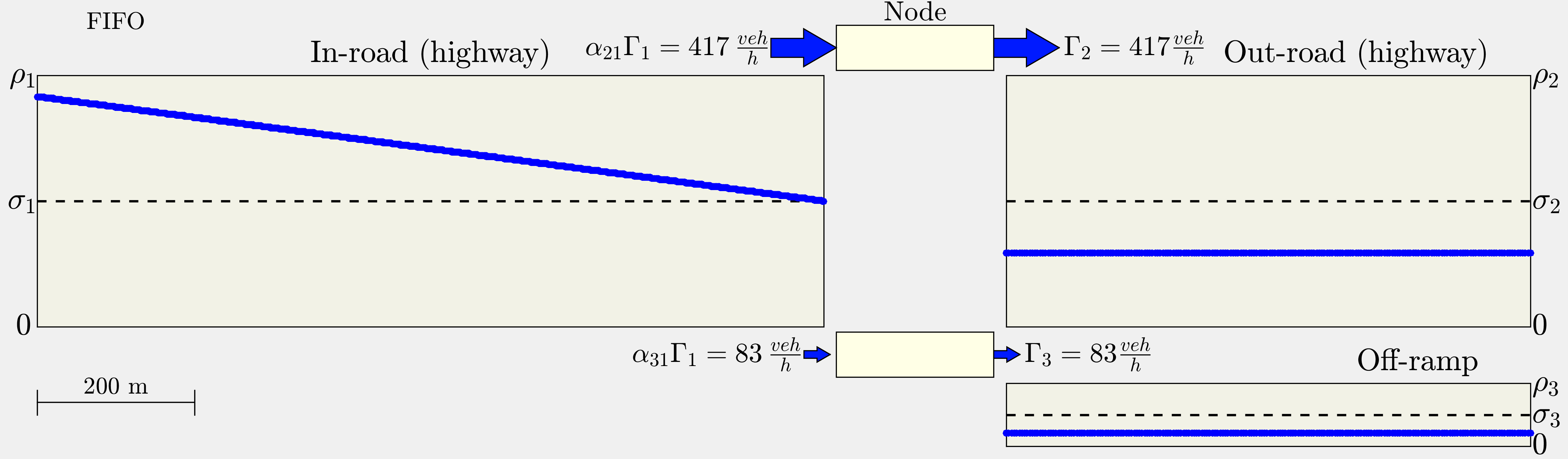}}}\\[-.1em]
\subfigure[$t=25$ min]{\parbox{\subfigwidth}{\label{fig:FIFO_t_25}\includegraphics[width=\subfigwidth]{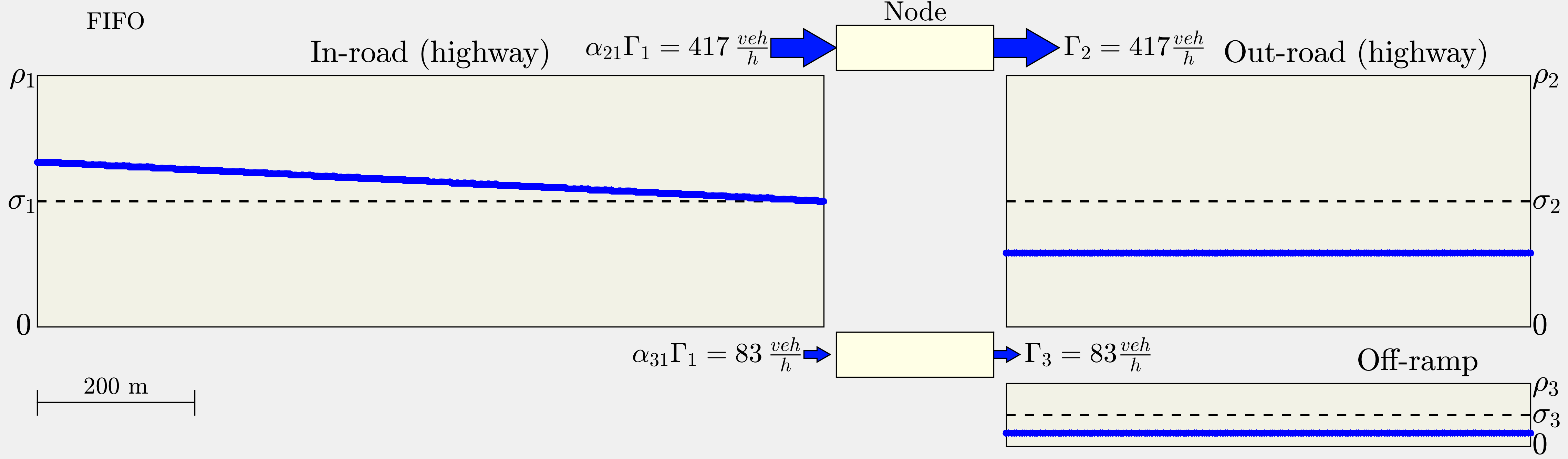}}}
\caption{Time evolution of the solution of the FIFO model.}\label{fig:FIFO}
\end{figure}

\begin{figure}[p]
\centering
\subfigure[$t=1.5$ min]{\parbox{\subfigwidth}{\label{fig:NONFIFO_t_1_5}\includegraphics[width=\subfigwidth]{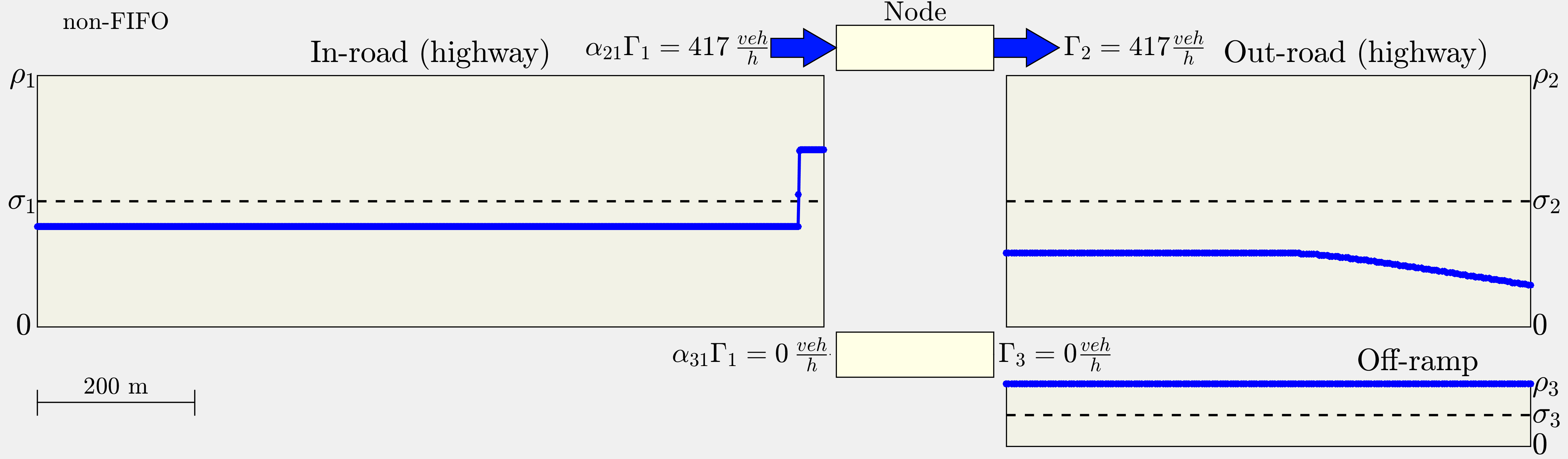}}}\\[-.1em]
\subfigure[$t=6$ min]{\parbox{\subfigwidth}{\label{fig:NONFIFO_t_6}\includegraphics[width=\subfigwidth]{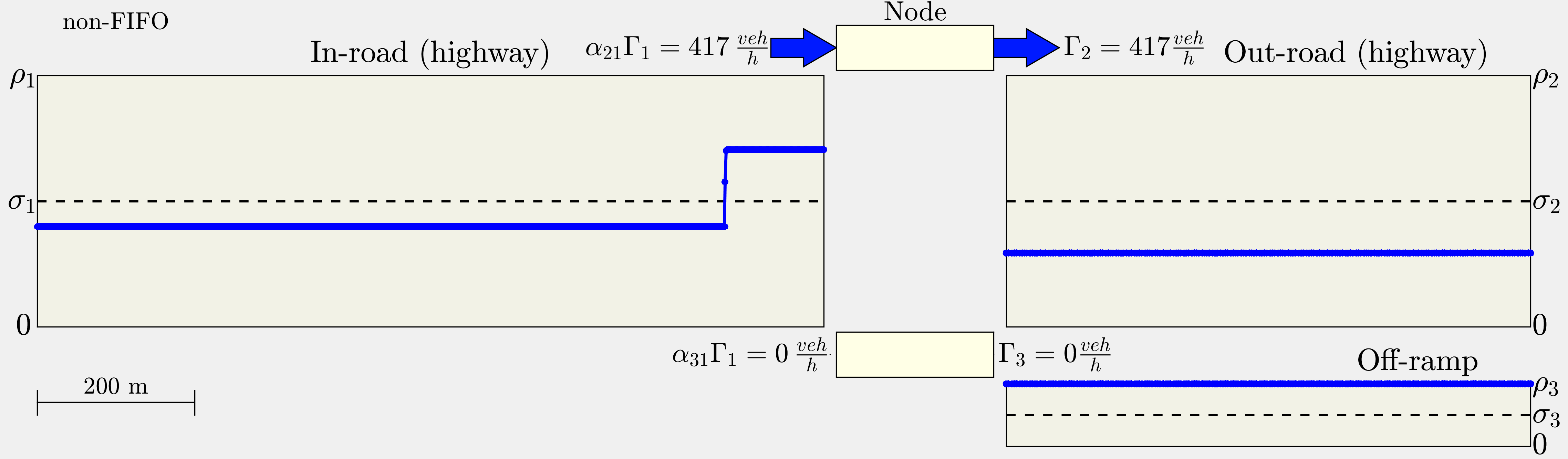}}}\\[-.1em]
\subfigure[$t=9.5$ min]{\parbox{\subfigwidth}{\label{fig:NONFIFO_t_9_5}\includegraphics[width=\subfigwidth]{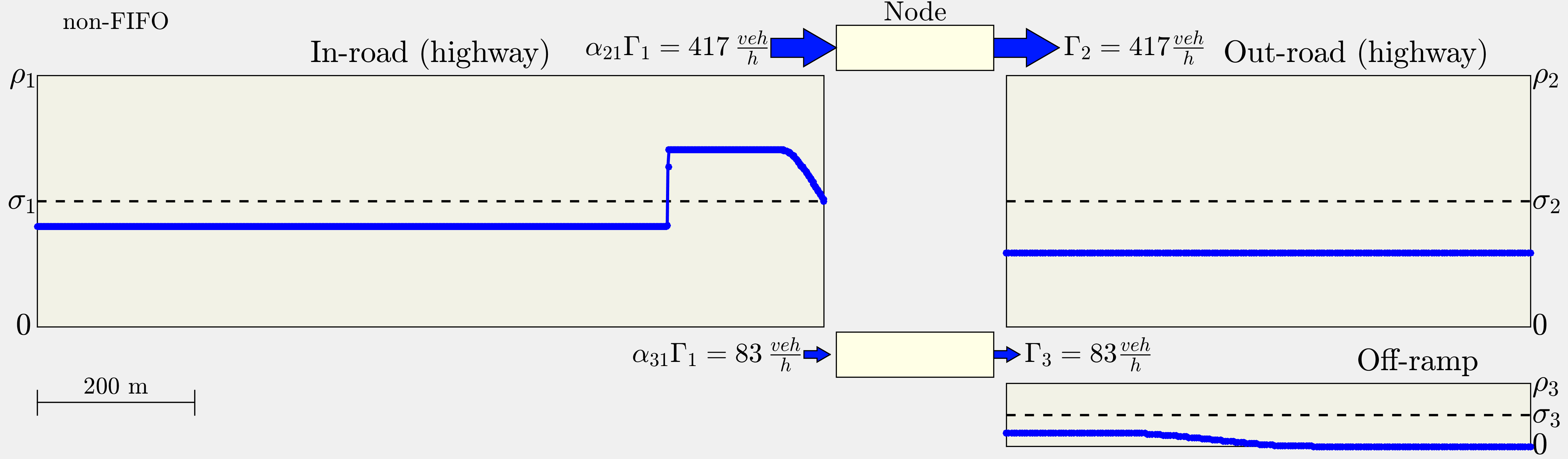}}}\\[-.1em]
\subfigure[$t=15$ min]{\parbox{\subfigwidth}{\label{fig:NONFIFO_t_15}\includegraphics[width=\subfigwidth]{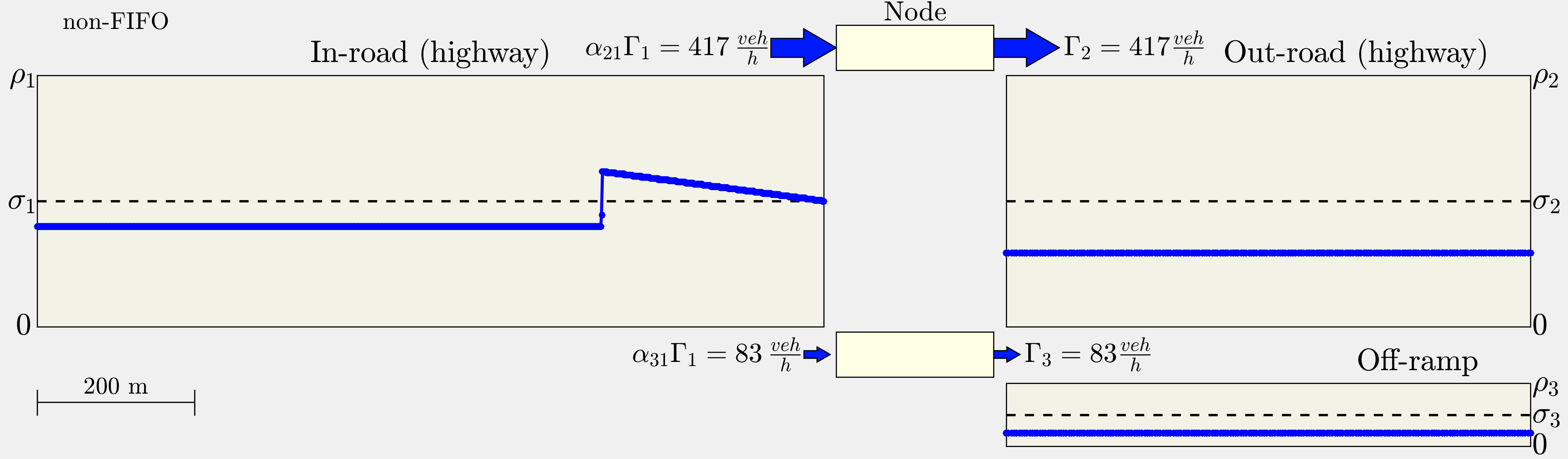}}}\\[-.1em]
\subfigure[$t=25$ min]{\parbox{\subfigwidth}{\label{fig:NONFIFO_t_25}\includegraphics[width=\subfigwidth]{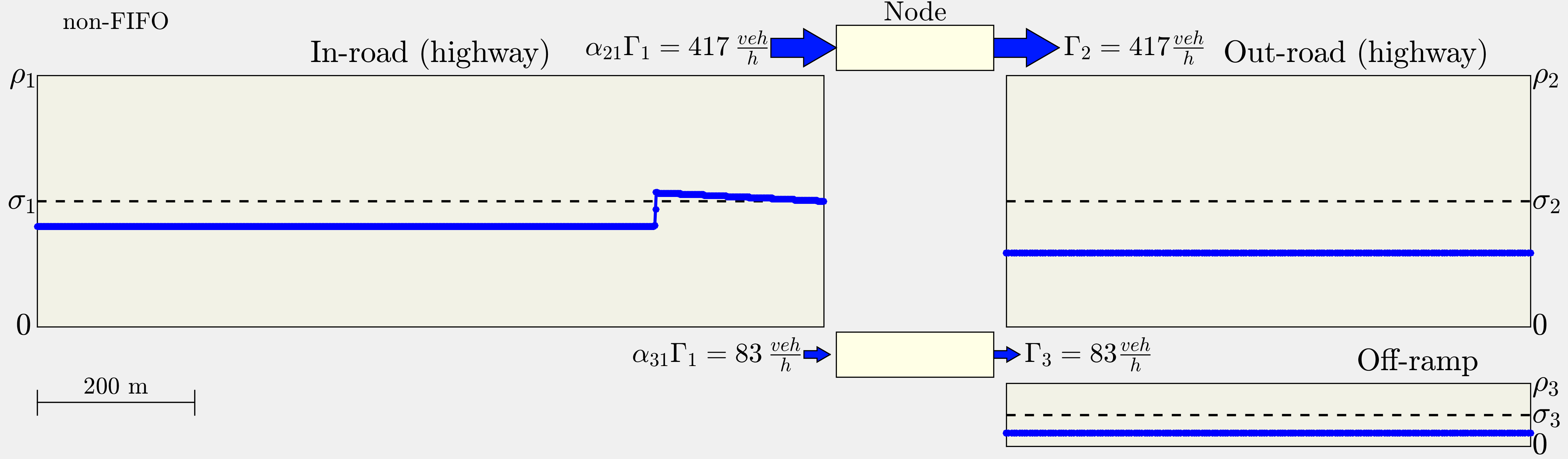}}}
\caption{Time evolution of the solution of the non-FIFO model.}\label{fig:NONFIFO}
\end{figure}

\begin{figure}[p]
\centering
\subfigure[$t=1.5$ min]{\parbox{\subfigwidth}{\label{fig:FIFOWQueue_t_1_5}\includegraphics[width=\subfigwidth]{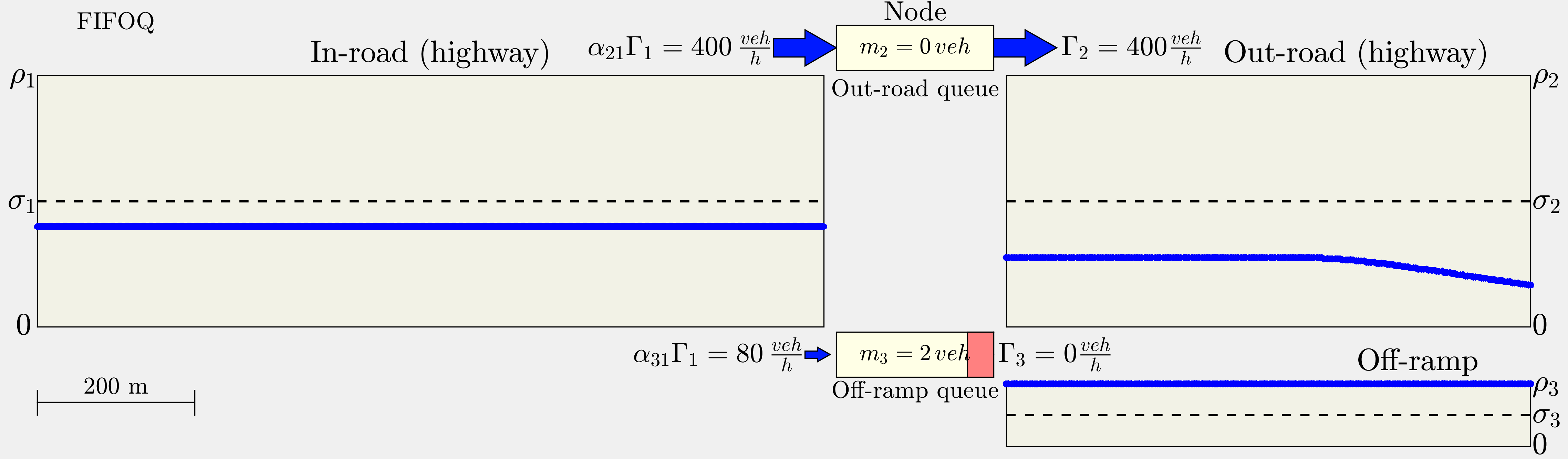}}}\\[-.1em]
\subfigure[$t=6$ min]{\parbox{\subfigwidth}{\label{fig:FIFOWQueue_t_6}\includegraphics[width=\subfigwidth]{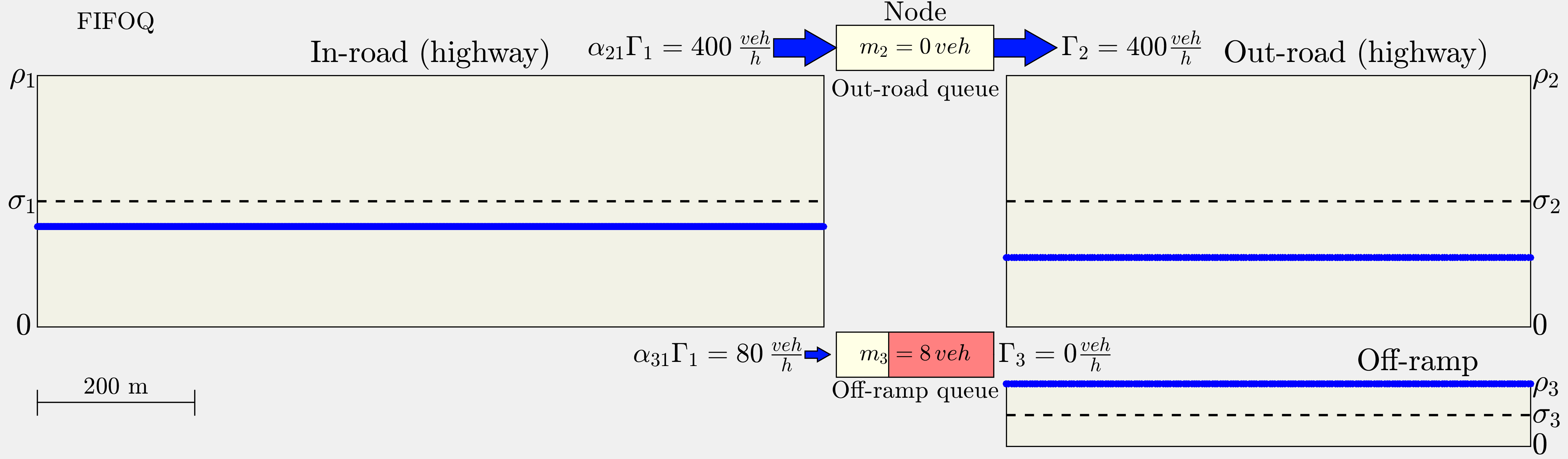}}}\\[-.1em]
\subfigure[$t=9.5$ min]{\parbox{\subfigwidth}{\label{fig:FIFOWQueue_t_9_5}\includegraphics[width=\subfigwidth]{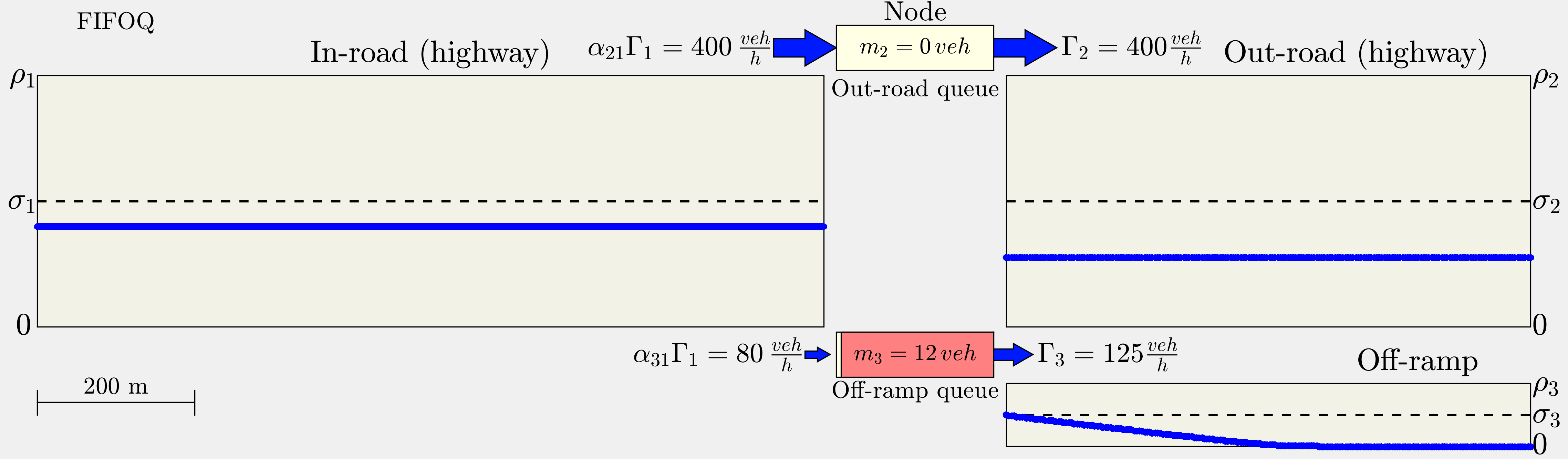}}}\\[-.1em]
\subfigure[$t=15$ min]{\parbox{\subfigwidth}{\label{fig:FIFOWQueue_t_15}\includegraphics[width=\subfigwidth]{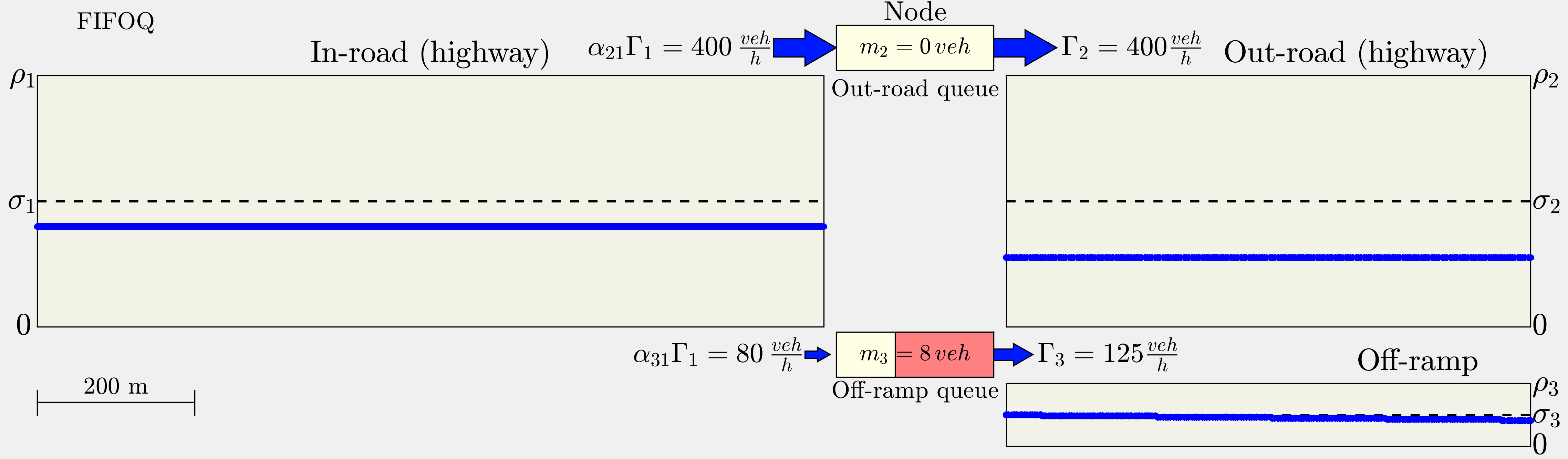}}}\\[-.1em]
\subfigure[$t=25$ min]{\parbox{\subfigwidth}{\label{fig:FIFOWQueue_t_25}\includegraphics[width=\subfigwidth]{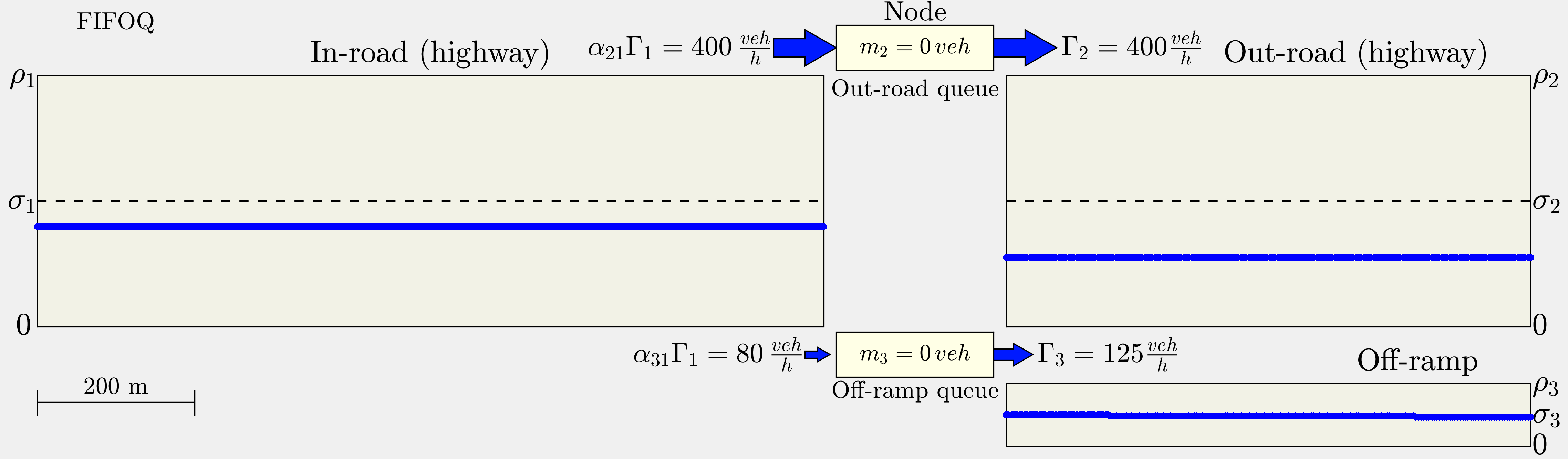}}}
\caption{Time evolution of the solution of the FIFOQ model.}\label{fig:FIFOWQueue}
\end{figure}

Figures~\ref{fig:FIFO}, \ref{fig:NONFIFO}, and~\ref{fig:FIFOWQueue} show the time evolution (at five representative snapshots in time) of the solutions produced by the three models, computed by a highly resolved Godunov/CTM discretization (with proper queue treatment). For the FIFOQ model, Figure~\ref{fig:FIFOWQueue} shows the magnitude of the queue $m_3$ in the lower queue box. The blue arrows visualize the flows into and out of the respective queues. For consistency, empty queue boxes are also shown for the models without queues.

For both classical models, Figures~\ref{fig:FIFO_t_1_5} and~\ref{fig:FIFO_t_6}, as well as Figures~\ref{fig:NONFIFO_t_1_5} and~\ref{fig:NONFIFO_t_6}, show the initial backwards propagating shock that arises due to the clogged off-ramp. Clearly, with FIFO, the level of congestion on the in-road is much larger (in fact, completely jammed) than with non-FIFO. For the FIFOQ model (Figures~\ref{fig:FIFOWQueue_t_1_5} and~\ref{fig:FIFOWQueue_t_6}) the accumulation of type 3 vehicles is instead tracked in the queue.

Then, for all models, Figures~\ref{fig:FIFO_t_9_5}, \ref{fig:NONFIFO_t_9_5}, and~\ref{fig:FIFOWQueue_t_9_5} show the system state right after the clearing of the off-ramp (the idealized maneuver). For both FIFO and non-FIFO, the flow turns maximal, resulting in a rarefaction fan on the in-road. In turn, in the FIFOQ model the queue $m_3$ has started to decrease.

Finally, for both FIFO and non-FIFO, Figures~\ref{fig:FIFO_t_15} and~\ref{fig:FIFO_t_25}, as well as Figures~\ref{fig:NONFIFO_t_15} and~\ref{fig:NONFIFO_t_25}, show the gradual approach of the system towards a uniform state on the in-road. In contrast, for FIFOQ the in-road has remained in free-flow the whole time. Figures~\ref{fig:FIFOWQueue_t_15} and~\ref{fig:FIFOWQueue_t_25} show the shrinking and eventual depletion of the queue. Right after $t=25$ min, the outflow states will change to what the other two models would also yield as $t\to\infty$.

\vspace{1.5em}
\section{Conclusions and Outlook}\label{Conclusions}
In this paper, we have introduced a new coupling model for macroscopic lane-aggregated traffic flow at 1-in-2-out nodes modeling off-ramps: FIFOQ. The new model demonstratively remedies fundamental modeling shortcomings of the existing FIFO and non-FIFO models (blockage and spurious re-routing, respectively). It achieves this goal by means of a vertical queue that tracks the excess of vehicles of a certain type. A cell transmission discretization of the new model has been presented, and applied in a representative example.

The new model is clearly devoid of the fundamental shortcomings of existing models. Of course that does not necessarily imply that the new model is descriptive of reality; all we know is that it is ``less bad'' than existing models. A model validation with real traffic data will be an important step for future research.

On the modeling side, one key simplification assumed in this paper is that upstream of the off-ramp, traffic divides the width of the highway according to the split ratio. This is clearly unrealistic in most situations, where off-ramp queues commonly are restricted to the right-most lane. Future work will be devoted to extending the model to the more general situation. In fact, on real highways the road width available for passing an off-ramp queue does depend on the queue length itself, as long queues tend to affect other lanes than merely the right-most lane. Again, such modeling extensions will be explored in the future.

\vspace{1.5em}
\section*{Acknowledgments}
B.~Seibold would like to acknowledge support by the National Science Foundation through grant CNS--1446690.

\vspace{1.5em}
\bibliographystyle{plain}
\bibliography{references_complete}

\vspace{1.5em}
\end{document}